\documentclass[aps,twocolumn,superscriptaddress,prb,amsmath,amssymb,showpacs,lengthcheck,reprint,floatfix]{revtex4-1} 

\usepackage[T1]{fontenc}

\usepackage[latin9]{inputenc}

\usepackage{mathrsfs}

\usepackage{graphicx}

\usepackage{color}

     \newcommand{\note}[1]{{#1}}

\usepackage{braket}

\usepackage{placeins}

\begin{document}

\title{A Koopmans-compliant screened exchange potential with correct asymptotic behavior for semiconductors}  
\author{Michael Lorke}
\affiliation{Institute for Theoretical Physics, University of Bremen, Otto-Hahn-Allee 1, 28359 Bremen, Germany}
\affiliation{Bremen Center for Computational Materials Science, University of Bremen, Am Fallturm 1, 28359 Bremen, Germany}
\author{Peter De\'{a}k}
\affiliation{Bremen Center for Computational Materials Science, University of Bremen, Am Fallturm 1, 28359 Bremen, Germany}
\author{Thomas Frauenheim}
\affiliation{Bremen Center for Computational Materials Science, University of Bremen, Am Fallturm 1, 28359 Bremen, Germany}
\affiliation{Computational Science Research Center, No.10 East Xibeiwang Road, Haidian District, Beijing 100193}
\affiliation{Computational Science and Applied Research Institute Shenzhen, China.}
\begin{abstract}

The performance of density functional theory depends largely on the approximation applied for the exchange functional.
We propose here a novel screened exchange potential for semiconductors, with parameters based on the
physical properties of the underlying microscopic screening and obeying the requirements for proper asymptotic behavior. 
We demonstrate that this functional is Koopmans-compliant and reproduces
a wide range of band gaps. We also show, that the only tunable parameter of the functional can be kept constant upon changing  
the cation or the anion isovalently, making the approach suitable for treating alloys.

\end{abstract}

\maketitle

%

Density functional theory (DFT) is the workhorse of electronic structure calculations in many areas of solid state physics 
and materials science. 
The exact form of the exchange-correlation potential is, unfortunately, not known, and the real predictive power of 
DFT-based methods is mainly limited by the quality of the applied approximate exchange functional. 
It has been established \cite{Perdew:82} that the exact functional provides a total energy, which is a 
piecewise linear function of the occupation numbers and has
a derivative discontinuity at integer values. 
This is not case for the standard approximations (LDA: local density approximation, or GGA: generalized gradient approximation). 
As it is well known, these functionals are convex between, and have no derivative discontinuity at integer occupation numbers. 
Therefore, they underestimate the gap of semiconductors and lead to an artificial delocalization of defect states. 
In contrast, DFT with unscreened non-local Hartree-Fock (HF) type exchange 
leads to a strong overestimation of the band gap and to over-localization of defect states. 
Earlier, self-interaction correction schemes to LDA/GGA have been applied to remedy these problems \cite{Perdew:81,Baumeier:06}, but in the last decade, 
screened exchange \cite{Clark:10,Clark:10a} and hybrid 
functionals \cite{Becke:96,PBE0,HSE,Krukau:08,Zheng:11,Alkauskas:11,Skone:14,Chen:18} (that mix semi-local 
and non-local exchange) have emerged as useful alternatives.

Hybrid functionals utilize error compensation between semi-local and non-local exchange \cite{Deak:19a} 
and their parameters are tuned semi-empirically. While connection between the parameters and the screening 
properties of the solid were being sought \cite{Marques:11,Skone:14,Chen:18}, 
the importance of Koopmans-compliance became ever more 
recognized \cite{Stein:10,Miceli:18,Nguyen:18}. 
Fulfillment of the generalized Koopmans' theorem (gKT) is equivalent to the linearity of the total energy as a 
function of the fractional occupation numbers,\cite{Janak:78,Perdew:87,Perdew:97a} and is the precondition for 
predicting the correct localization of one-electron states\cite{Lany:09}. So-called Koopman's compliant 
functionals, (see, e.g., Refs.~\cite{Perdew:81SI,Dabo:10,Nguyen:18}), fulfill the gKT by construction, 
which is appealing from a fundamental viewpoint. \note{However, at present their  
applicability for solids suffers from practical limitations due to, e.g., limited possibility for k-point sampling.}

Earlier we have shown that the two parameters of the Heyd-Scuseria-Ernzernhof (HSE) hybrid \cite{HSE} 
can be tuned to reproduce the band gap and fulfill the gKT \cite{Deak:19a}. 
This has allowed to obtain very accurate results for defects, but 
the parameters had to be optimized for every material separately \cite{Deak:10,Han:17,Deak:17,Deak:19}.

Current hybrid functionals, used for calculating defects in solids, like PBE0  \cite{PBE0} and HSE \cite{HSE} but also 
the sX (screened exchange) \cite{Clark:10} functional, do not show the physically 
correct asymptotic behavior of the screening in semiconductors.
As an alternative, we present here a novel screened exchange functional with proper asymptotic behavior. 
We will show that this functional reproduces the relative 
position of the band edge states and fulfills the gKT, i.e., 
the total energy shows the correct piecewise linearity 
as the function of the occupation numbers. 
These properties make the functional very well suited for defect calculations in semiconductors.
The parameters of this functional are derived from the physical properties of the screening, 
and it contains only one adjustable parameter with a value being constant upon substitution 
of the cation or the anion in a given class of semiconductors. (\note{As we will show, even that 
parameter can be obtained approximatively from first-principles.})

We start from the ansatz
\begin{equation}
V^\text{X}(q)=\varepsilon^{-1}(q)V_\text{HF}^\text{x}(q) ~.
\end{equation}
Here $V_\text{HF}^\text{x}(q)$ is the Hartree-Fock-type non-local exchange potential in wave vector space,
and the model screening function is
\begin{equation}\label{eps1}
\varepsilon^{-1}(q)=1+\left(\frac{1}{\varepsilon_{b}}-1\right) \frac{1}{\cosh(q/\sigma)}~.
\end{equation}
Correlations are added on the GGA level in the PBE (Perdew, Burke, Ernznerhof) approximation \cite{PBE}.
The choice of an $1/\cosh$ behavior is guided by a well known result of Green's function theory,
that the $q\rightarrow\infty$ behavior should be exponential \cite{Banyai:98,Gartner:02} while the 
$q\rightarrow~0$ behavior has to be quadratic \cite{Gartner:00}.
Our choice  for $\varepsilon$ ensures that $V^\text{X}(r)$ has the 
proper 1/r behavior for $\vec{r}\rightarrow \infty$  and 
at the same time approaches the correct pure Coulomb limit at $\vec{r}\rightarrow  0$.

\begin{figure}[!ht]
\centering
\includegraphics[width=\columnwidth,angle=0]{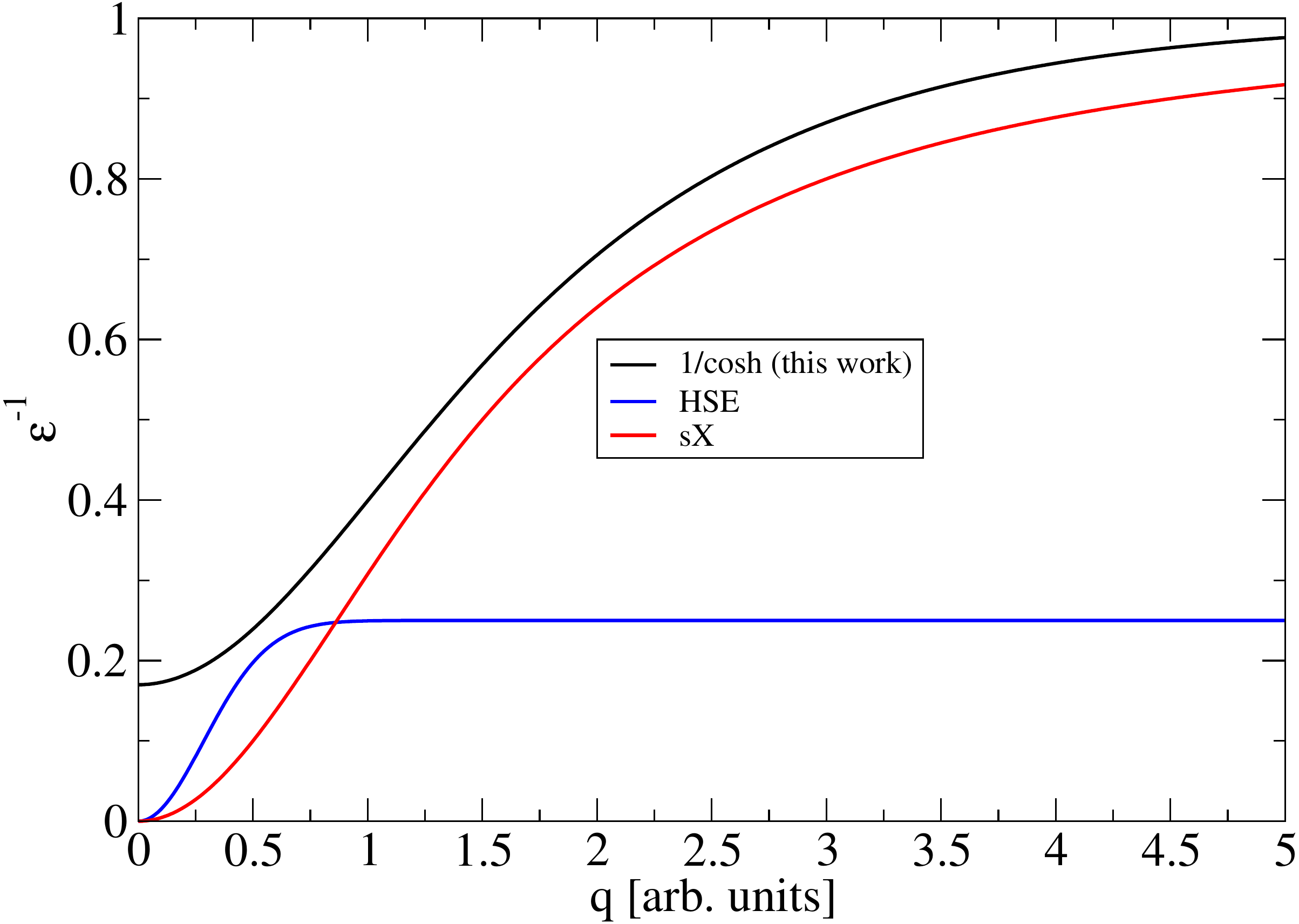}
\caption{(a) Screening functions $\varepsilon^{-1}(q)$ as defined in Eqs.~\eqref{eps1} for GaN. 
For comparison the HSE hybrid and the sX functional are shown. 
\label{fig1}}
\end{figure}

\note{To determine the screening length $\sigma$, we follow Refs.~\cite{Cappellini:93,Shimazaki:08}
in deriving a static approximation to the dielectric function in random phase approximation (RPA).
This results in a Lorentzian dependence of $\varepsilon^{-1}(q)$ with a screening length $\kappa$.
We than choose the same FWHM in our $\cosh$-ansatz as the RPA would give for its Lorentzian screening.
This procedure results in }
\begin{equation}\label{eq:screening_length}
 \sigma=\frac{Z}{  \log(2+\sqrt{3})} \sqrt{k^2_{\rm TF} \left( \frac{1}{\varepsilon_{b}-1} + 1 \right)}~.
\end{equation}
The TF wave vector $k_{\rm TF}$ can be expressed as \cite{Cappellini:93,Shimazaki:08,Shimazaki:10}
\begin{equation}\label{eq:screening_k}
k_{\rm TF}=4\left(\frac{3 N_{\rm el}}{ \pi V}\right)^{1/3}~, 
\end{equation}
with the cell volume $V$.
We will elaborate on the choice of the effective number of electrons $N_{\rm el}$ per unit cell and on the renormalization factor $Z$ below.
Our exchange potential can be seen as a static approximation to a  $GW$ calculation with a model screening function. 
In it's idea, the method is similar to the sX functional \cite{Clark:10}, however, 
the inclusion of the correct limits and q-dependence marks a substantial improvement.
In Fig.~\ref{fig1}, we compare our model screening function to the ones used in the HSE and sX functionals.
The sX clearly neglects the background screening $\varepsilon_{b}$ and uses a Lorentzian q-dependence, while
the HSE approach differs substantially and compensates by mixing with PBE exchange. 

To apply this theoretical model to semiconductor materials we have to determine the dielectric constant
$\varepsilon_{b}$, the effective number of electrons $N_{\rm el}$ and the renormalization factor $Z$.
The value of  $\varepsilon_{b}$ is taken from an independent particle calculation of the optical properties at 
the GGA (PBE) level. 
This is in the spirit of the GW$_0$ approximation \cite{Shishkin:07}, which in most cases 
reproduces the band gaps better than fully self-consistent GW schemes (without vertex corrections). 

It is well known, that energetically deep lying electronic states contribute insignificantly to the screening, 
as they are strongly localized, 
their spatial overlap is weak and their energetic distance to the top of the VB is high. 
Therefore, as an approximation, we disregard them in the screening completely.
The ''effective number of electrons'', $N_{\rm el}$, is hence determined as follows. We evaluate which states 
primarily form the highest valence band and choose the corresponding 
number of electrons from the constituting atomic states 
\footnote{technically from \textsc{Vasp}'s POTCAR file}. 
As an example, in GaN, the top of the VB is made up of nitrogen 2p orbitals. In a 
nitrogen atom, these hold 3 electrons, hence for the primitive cell, containing 2 
nitrogen atoms, $N_{\rm el}=6$. $N_{\rm el}$ should, of course, be increased for a supercell, 
according to the number of atoms involved.
It should be noted, that neither $Z$ nor $N_{\rm el}$ 
enter $\sigma$ independently, but only in combination $ZN_{\rm el}^{1/3}$.  Therefore  the choices for $N_{\rm el}$ and $Z$ are not unique, as they lead to the same functional  
if the same  $\sigma$ is generated.

To understand the role of $Z$, let us recall that in $GW$ theory the screened potential $W$ is usually 
calculated in RPA. The polarizability $P$ that controls $W$ via the 
Dyson equation $W = V + VPW$ (V is the unscreened Coulomb potential), 
is given by the product of two Green's functions, $P = GG$  \cite{Baym:62}. 
We base our approach on the same framework, with the role of $W$ taken by $V_X$. 
One can define a renormalization factor $Z$ that describes the fraction of spectral weight attributed 
to the main quasi-particle peak within the full Green's function (GF) $G$. In our approximation only 
this fraction $Z$ of the full GF $G$ is considered in determining the polarizability $P$. 
Via the RPA ($P = GG$ ), 
this leads (with the usual steps used to derive static screening \cite{Schaefer:02}) to the occurence 
of $Z$ in Eq.~\eqref{eq:screening_length}. 
Physically, this means that we assume significant contributions to the screening only by the 
main quasi-particle peak of the GF G.

Obviously, this value of $Z$ varies for the states around the valence band edge 
and between different bands $\lambda$ in the $GW$ approximation,
i.e. $Z_\text{GW}$=$Z_{\lambda}(\vec{k})$. As an approximation we average
Z over the bands that hold the $N_{\rm el}$ electrons to get an average value $\overline{Z_\text{GW}}$.
However, as  effects
beyond a static approximation and beyond the RPA clearly influence the screening, 
we employ a pragmatic approach, 
using $\overline{Z_\text{GW}}$ as a starting point, and modifying it to reproduce the band gap,
yielding Z$_\text{opt}$. As we will show below, both
the average value $\overline{Z_\text{GW}}$ and hence also the optimal value Z$_\text{opt}$
are constant for isovalent substitution of atoms.This allows 
to tackle semiconductor alloys with a common $Z$ value.


We have implemented the exchange functional as discussed above into the Vienna Ab initio Simulation Package, \textsc{Vasp} 5.4.4 \cite{VASP:3,VASP:4}, using the projector augmented wave method and treating the semi-core d-states as part of the valence shell. 
The modified \textsc{Vasp} source code can be made available to certified owners of a \textsc{Vasp} 
user license. 
Calculations on the unit cell of the bulk materials where performed 
using a 6$\times$6$\times$6 $\Gamma$-centered 
Monkhorst-Pack \cite{Monkhorst:76} grid. For defect 
calculations, 64 (GaAs), 96 (GaN), 160 (Ga$_2$O$_3$), and 512 (diamond) atom supercells were used, 
applying the $\Gamma$-point approximation.
The defect geometries were fully relaxed.
A 450eV (900eV) cutoff was applied for the expansion of the wave functions (charge density).
The computational cost for practical calculations of defects and adsobates is on par with that of DFT calculations
with the PBE0 exchange potential.
Charge corrections for the total energy were performed by the method of Freysoldt, Neugebauer, and van der Walle \cite{Freysoldt:09},
while localized defect levels were corrected using the formula derived by Chen and Pasquarello \cite{Chen:13}. 
GW$_0$ calculations were performed on top of PBE calculations with 1000 bands 
and a 6$\times$6$\times$6 $\Gamma$-centered Monkhorst-Pack grid to provide a reference for the band 
gap and a starting value for $Z$. 
In the present calculations, the experimental lattice constants are
used for consistency with the literature.

For the purpose of this letter, we study 13 semiconductor compounds with band gaps ranging between 0.5 and 14 eV. 
The band gap of most of these materials is systematically underestimated by standard HSE06\cite{Deak:19a}.  
These compounds represent different lattice 
structures and vary in composition and in the degree of ionicity, therefore, they can serve as a representative 
set for testing the method.

\begin{figure}[h]
\centering
\includegraphics[width=\columnwidth,angle=0]{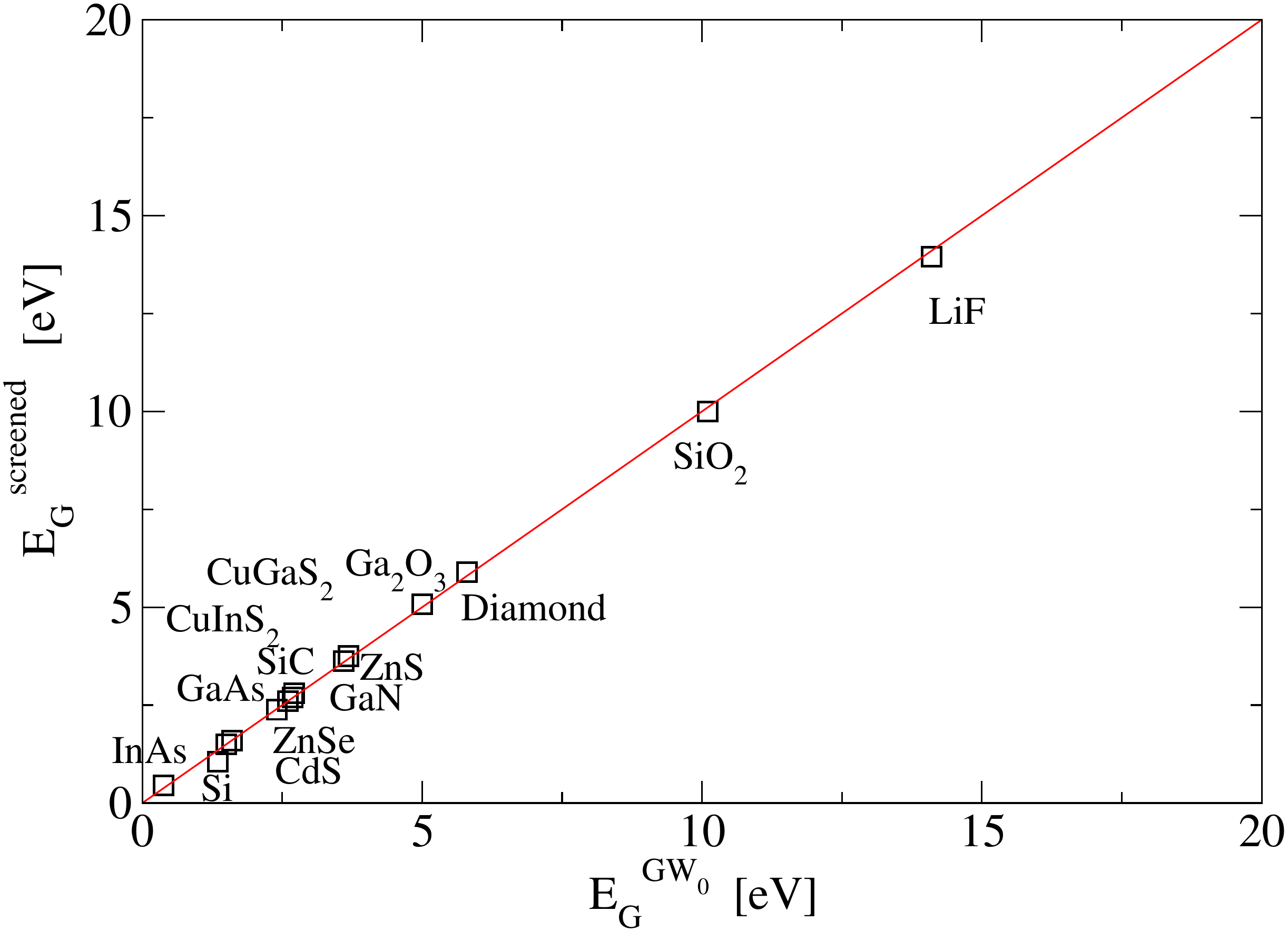}
\caption{Fundamental band gap E$_G^\text{screened}$ for a variety of semiconductors, as calculated via the screened exchange approach proposed. Depicted is the band gap 
resulting of our functional as a function of the GW$_0$ or experimental band gap E$_G^{GW_0}$.
\label{fig:bands}}
\end{figure}

Table~\ref{table:gap} shows the calculated band gaps, in comparison with the GW$_0$ results. The 
optimal value of the scaling factor Z$_\text{opt}$ is also given.
We note that our value for the quasi-particle band gap, as given in Table~\ref{table:gap}, is determined 
to include neither excitonic nor polaronic renormalizations.
Both the electron-hole Coulomb interaction and the  electron-phonon interaction lower 
the band gap (resulting in the measurable optical gap) and 
are not included here \cite{Deak:17,Deak:19}. An overview of the comparison between the band gaps 
of the proposed method and the GW$_0$ ones is given in Fig.~\ref{fig:bands}.
We find excellent agreement, with a mean average error of 0.05eV and a mean average percentage error 
of only 3\%.

\begin{table}[h]
 \begin{center}   
\begin{tabular}[t]{l|c|c|c|c|c|c|c}
  & \phantom{}E$_G^{\text{target}}$ (GW$_0$)   & \phantom{ss}Z$_\text{opt}$\phantom{ss} &\phantom{}E$_G$ (present work)  \\\hline
  LiF & 14.2  &0.62 &     14.2                                                                           \\\hline
  ZnSe& 2.7 &0.62 &     2.8                                                                                       \\
  ZnS & 3.7 &0.62 &     3.8                                                                              \\
  CdS & 2.4  &0.62 &     2.4                                                                             \\\hline
  InAs& 0.4  &0.72 & 0.4                                                                               \\
GaN & 3.6 &0.72& 3.6                                                                                    \\\hline
Ga$_2$O$_3$& 5.0 & 0.72 &5.1                                                                             \\\hline
CuGaS$_2$ & 2.6  &0.65& 2.6                                                                              \\
CuInS$_2$ & 1.5 &0.65& 1.5                                                                                \\\hline
Si &1.2 & 0.53 & 1.1                                                                                    \\
3C-SiC & 2.7&0.53& 2.7                                                                                  \\
Diamond & 5.8 & 0.53 & 5.8                                                                               \\
SiO$_2$ & 10.2 & 0.53 & 10.1                                                                              \\\hline
\end{tabular}
\caption{Fundamental 0K quasi-particle band gap of various materials. Given are the renormalization factor $Z$, and the 
resulting band gap with the screened exchange functional presented in this work.
}
\label{table:gap}
\end{center}
\end{table}

 A striking feature of these results is that, in a particular sub-class of 
 materials, the optimal value Z$_\text{opt}$  remains the same when the cation or the anion is replaced.
The reason for that is a similar behavior of the average  $GW_0$ value, $\overline{Z_\text{GW}}$,
as shown in Table \ref{table:z}. We can infer that Z$_\text{opt}$
remains constant for all II/VI, III/V, and IV/IV compounds. 
Deviations  between the optimal and average $GW$ values most probably stems from the use 
of a static approximation to the GW self energy in Eq.~\eqref{eps1} and hence 
they reflect the influence of frequency dependent screening. 
The biggest deviation is found for silicon (1.23eV vs.~1.05eV),
where it is known \cite{Thygesen:13} that static approximations tend to underestimate the band gap. 
This is also the case for our approximation, 
where the agreement for silicon is worse than for SiC or diamond.

The  results above mean that our method allows to treat alloys of isovalent 
elements without parameter retuning. It is important to note that, as we will discuss below, 
not only the band gap is reproduced, 
independent of the alloy composition, but also the gKT is satisfied.

\begin{table}[h]
 \begin{center}   
\begin{tabular}[t]{l|c|c|c|c|c|}
	                  & \phantom{ss}GaN\phantom{ss}   &\phantom{ss}ZnS \phantom{ss}&\phantom{ss} CdS\phantom{ss} &\phantom{ss} LiF\phantom{ss} &diamond \\\hline
  $\overline{Z_\text{GW}}$& 0.80 &0.77& 0.77&   0.77&  0.81        \\\hline
 Z$_\text{opt}$          & 0.72 &0.62 &  0.62 & 0.62&0.53	\\\hline
  \end{tabular}
\caption{Z values from the $GW$ calculation, averaged over the states filled by the $N_{\rm el}$ electrons, for different materials
and in comparison to the optimal values Z$_\text{opt}$ used in Eq.~\eqref{eq:screening_length}
}
  \label{table:z}
\end{center}
\end{table}

We want to point out, that, in an alloy, $\varepsilon_b$ has to be 
recalculated for the given composition, while $N_{\rm el}$ is determined by the prevailing crystal structure.
In principle, the common Z value would allow to treat heterostructures between compounds of such sub classes 
but presently the spatial variation of the dielectric constant is not taken into account. Work in 
that direction is in progress.

\begin{figure}[h]
\centering
\includegraphics[width=\columnwidth,angle=0]{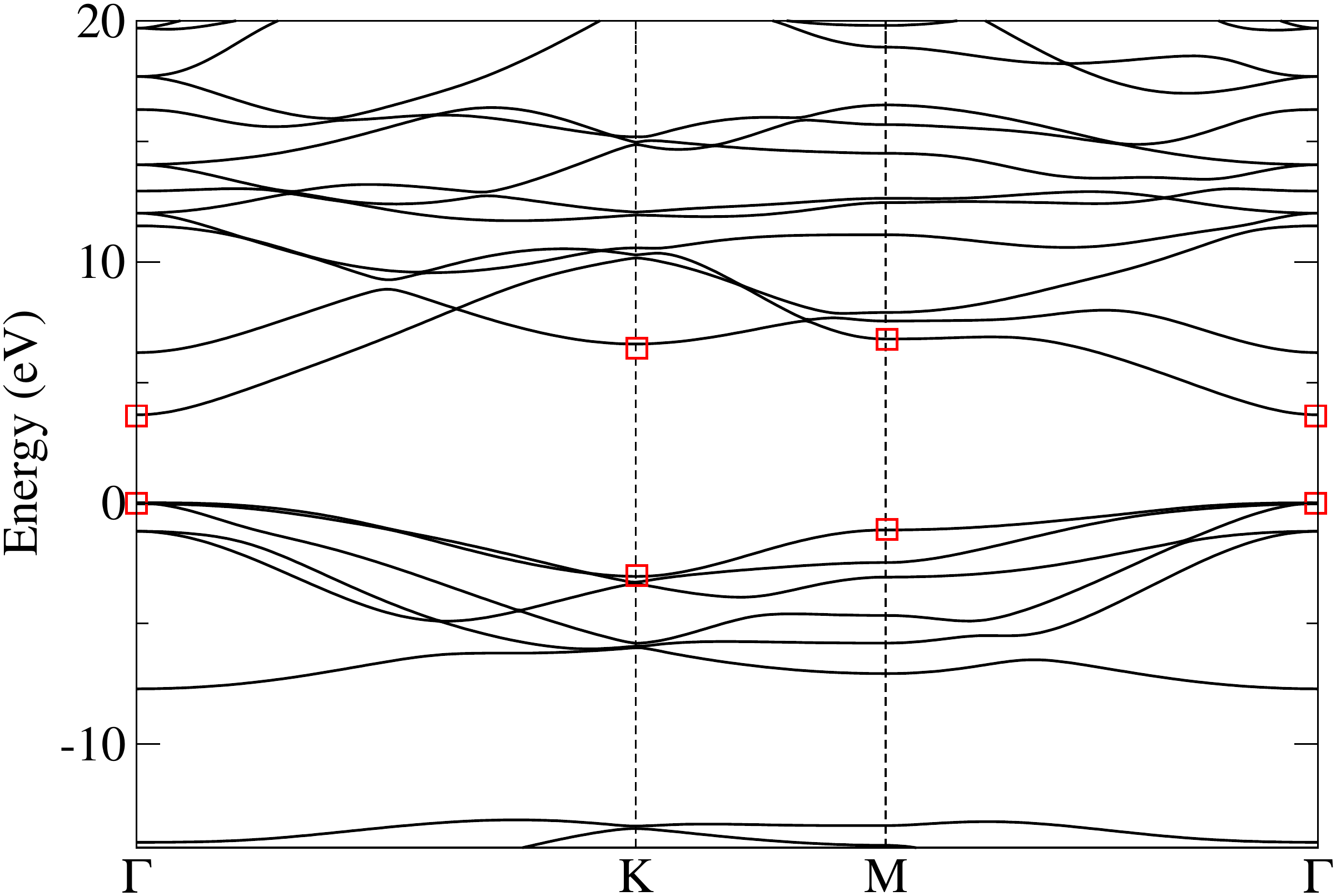}
\caption{Band structure of GaN, calculated with the GW$_0$ approach (lines) as well as with the screened exchange approach presented in this work (squares). 
The energy of the valence band maximum has been set to 0.
\label{fig2}}
\end{figure}
As discussed in previous works \cite{Deak:17,Deak:19}, not only the minimum band gap, but the 
band edges positions over the entire Brillouin-zone have to fit the results of a GW$_0$ calculation, in order to provide a proper description of 
defect levels, as they can be generated by a superposition of all band edge states. 
This condition is also satisfied with our approach, as shown on the example of GaN in Fig.~\ref{fig2}. 

\begin{table}[h]
 \begin{center}
\begin{tabular}[t]{l|c|c}
  & $\Delta$KS (HOMO)-$\Delta$SCF & $\Delta$SCF - $\Delta$KS (LUMO)\\\hline
GaAs & 0.02 & 0.03\\\hline
GaN & -0.04 & -0.04\\\hline
Ga$_2$O$_3$ & -0.03 &-0.04 \\\hline
Diamond & -0.05 & 0.05\\\hline
\end{tabular}
\caption{Fulfillment of the gKT for various materials. $\Delta$KS (HOMO) and $\Delta$KS (LUMO) describe the energetic position of 
the Kohn-Sham levels of the highest occupied molecular orbital for the neutral (HOMO), 
and of the lowest unoccupied molecular orbital (LUMO) for the +1 charge state of the defect.
$\Delta$SCF is the electron removal energy. For a more detailed explanation, see footnote [46].
}
\label{table:koopman}
\end{center}
\end{table} 

To correctly describe the energetic position and the localization of defect states, the 
gKT  has to be fulfilled,\cite{Deak:19a,Lany:09} i.e., the total energy has to be linear
with respect to the fractional occupation number\cite{Note2}.
We have tested this criterion on the (0/+1) transition levels of the 
antisite pair in GaAs, the nitrogen vacancy in GaN, the oxygen vacancy in Ga$_2$O$_3$, and the B$_\text{C}$ 
substitutional in diamond. As shown in Table~\ref{table:koopman}, we find an excellent fulfillment of the gKT.

It has been shown \cite{Marini:03} that issues in the description of optical properties within linear response TDDFT
(with the Cassida equation) are closely connected to the description of screening in the exchange kernel.
\begin{figure}[h]
\centering
\includegraphics[width=\columnwidth,angle=0]{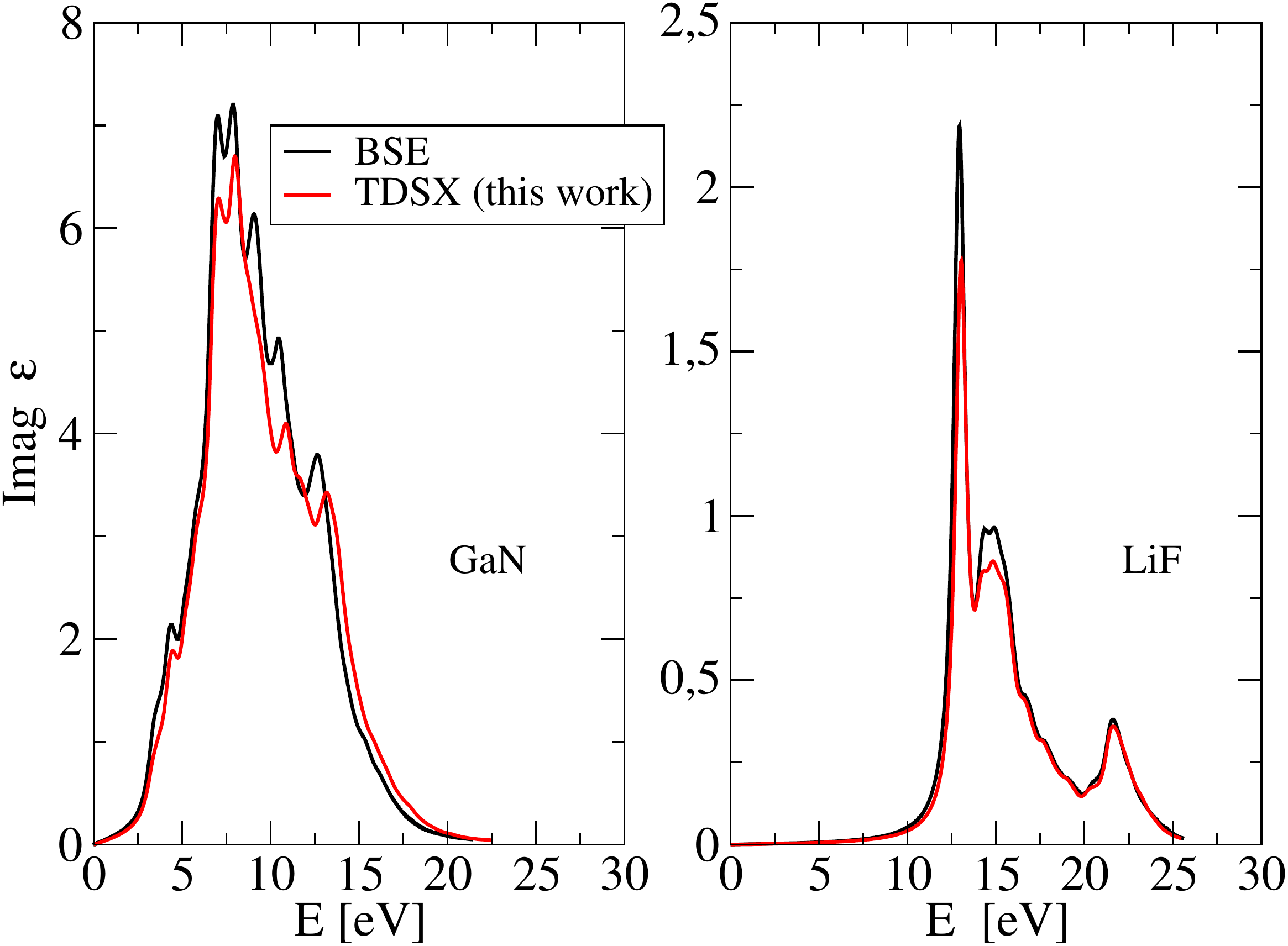}
\caption{
\label{fig:tddft} imaginary part of the optical response, calculated using the screened exchange method presented in this work (red) 
and the GW$_0$+BSE scheme (black).}
\end{figure}
Hence, we investigate the performance of our proposed method in comparison to state-of-the-art many-body perturbation theory
calculations with the GW$_0$+BSE scheme in Fig.~\ref{fig:tddft} for the cases of GaN and LiF with moderate and strong 
excitonic effects in the optical response, respectively. Apart from a slight underestimation of the oscillator strength,
very good agreement is found for a fraction of the computational effort.

In conclusion, we have presented a \note{functional based on exact exchange with improved screening} for semiconductors. 
As it is based on the correct asymptotic limits of the exchange potential, its parameters can be derived from physical principles,
to a large extent eliminating the need for ''tuning'' which is a common procedure with current hybrid functionals. 
Because of the correct piecewise linearity of the total energy, this functional reproduces the relative position of the band edge states 
and fulfills the generalized Koopmans' theorem. This is of high significance in various application areas. 
For example, in studying photo-assisted reactions on semiconducting catalysts, it is critical to reproduce the position of the band edges, 
which measure the chemical potential of photo-generated holes and electron \cite{Zhao:14,Deak:16}.
In general, the correct gap is the starting point to 
determine the optical and transport properties as well. 
Besides the reproduction of the gap, the fulfillment of the gKT is a condition
to accurately predict the localization and energy of defects states in semiconductors, which is a prerequisite in 
the successful identification 
of defects, which influence device behavior in micro/opto-electronics and photovoltaics \cite{Deak:19a,Deak:10,Deak:17,Deak:19}.
A great advantage of our approach is that it can be used to describe alloys without any re-tuning, as the single parameter 
of this approach is transferable between semiconductors of a similar type, when the cation or anion is replaced.
In addition, it is also suitable for the description of excitonic effects within linear response TDDFT, for a 
fraction of the cost of GW$_0$+BSE.

\begin{acknowledgments}

The authors would like to acknowledge stimulating discussions with G. Kresse. 
Funding from the DFG via research project FR2833/63-1 and the graduate school ``Quantum-Mechanical materials modeling'' 
as well as the support of the Supercomputer Center of Northern Germany via HLRN Grant No. hbc00027 is acknowledged.
\end{acknowledgments}

\appendix 
\section{Parameters}
In this appendix, we present computational details
about the calculation of the optical properties (MP sets and number of unoccupied states), 
and provide the effective number of electrons $N_{el}.$.
These results are summarized for all materials in Table~\ref{table:gapfull}.
It should be noted that the results for the dielectric constant
is strongly dependent on the number of bands, as also known from GW-type calculations.

\begin{table}[h]
 \begin{center}   
\begin{tabular}[t]{l|c|c|c|c}
        & NBands  & \phantom{xx} MP-set\phantom{xx} & \phantom{xx} $\varepsilon_b$\phantom{xx}  & \phantom{} N$_\text{el}$/unitcell  \\\hline
  ZnSe		&600&20$\times$ 20$\times$20& 8.29               & 4\\\hline
  ZnS 		&600&20$\times$ 20$\times$20& 6.35               & 4\\\hline
  CdS 		&600&20$\times$ 20$\times$20& 6.48               & 4\\\hline
  LiF 		&600&20$\times$ 20$\times$20&  2.03              & 5\\\hline
  InAs		&380&15$\times$ 15$\times$15&  19.8              & 3\\\hline
GaN 		&500&20$\times$ 20$\times$20& 5.82               & 6\\\hline
Ga$_2$O$_3$  	&500&20$\times$ 20$\times$20& 3.97               &24\\\hline
CuGaS$_2$ 	&300&15$\times$ 15$\times$15&  10.5              &16\\\hline
CuInS$_2$ 	&300&15$\times$ 15$\times$15&  10.8              &16\\\hline
Si 		& 60&12$\times$ 12$\times$12& 13.63              & 4\\\hline
3C-SiC 		& 60&12$\times$ 12$\times$12& 7.2                & 2\\\hline
Diamond 	& 60&12$\times$ 12$\times$12& 5.95               & 4\\\hline
SiO$_2$ 	&500&15$\times$ 15$\times$15& 2.46               &24\\\hline
\end{tabular}
\caption{Computational parameters used in the calculations.
Given are the number of bands, the MP set used in the PBE calculation of the dielectric constant,
the resulting dielectric constant $\varepsilon_{b}$, and the effective number of electrons per unitcell.
}
\label{table:gapfull}
\end{center}
\end{table}

\end{document}